\documentclass
{aa}
\usepackage{graphics}

\begin{document}
\thesaurus{11(11.11.1)}
\title{Order and chaos in galactic maps}
\author {N.D. Caranicolas and Ch.L. Vozikis}
\offprints{N.D. Caranicolas}
\mail{caranic@astro.auth.gr}
\institute{
Section of Astrophysics, Astronomy and Mechanics\\
Department of Physics,
University of Thessaloniki, GR-54006 Thessaloniki, Greece }
\date{Received ......, 1999 / Accepted ......, 1999} 
\titlerunning{Order and Chaos in Galactic Maps}
\authorrunning{N.D. Caranicolas and Ch.L. Vozikis}
\maketitle

\begin{abstract}                 
We investigate the properties of motion in a  map model derived from 
a galactic Hamiltonian  made up of perturbed elliptic oscillators. The
phase space portrait is obtained in all three different cases using
the map and numerical integration of the equations of motion. 
Our numerical calculations suggest that the map describes very well
the properties of regular motion in all three cases. There are cases
where the map fails to describe chaotic motion and cases where the map
describes satisfactorily  the chaotic phase plane. We also compare the  
Lyapunov characteristic number and the spectra of orbits  derived by
the map and numerical  integration  in each case. The agreement of the
outcomes is satisfactory. 
\end{abstract}


\section{Introduction}    
The application of modern methods in the domain of Galactic Dynamics
has been proved very fruitful  during the last decade. Among them is
the use of maps to describe galactic motion (Caranicolas \cite{c1},
\cite{c3}),
and the invariant spectra in galactic Hamiltonian systems (see Contopoulos
et al. \cite{cong}, Patsis et al. \cite{panos}).
 
Maps, derived from galactic type Hamiltonians, are useful in
the study of galactic orbits because they are faster, in general, than
numerical integration and allow a quick visualisation of the
corresponding phase plane. Maps also can give  the stability
conditions for the periodic orbits. On the other hand, our  experience
based on previous work shows that the results given by the maps are in
good agreement with those given by numerical integration at least for
small perturbations (see Caranicolas \& Karanis \cite{ck}). On this basis,
it seems that one has some good reasons to use a map for the study of
galactic motion.  

In the present work  we consider that the local (i.e. near an
equilibrium point ) galactic motion is described  by the potential
\begin{equation}
V=\frac{1}{2} \left( \omega_1^2 x^2 +  \omega_2^2 y^2 \right) - 
\epsilon \left[ \beta
\left(x^4 +y^4\right) + 2 \alpha x^2 y^2 \right] ~~,
\end{equation}
where $\omega_1, \omega_2$ are the unperturbed frequencies of oscillation
along the x and y axis respectively, $\epsilon >0$ is the perturbation
strength while $\alpha , \beta$ are parameters.We shall study the case
where $\omega_1= \omega_2 =\omega$. Without the loss of generality we
can take $\omega=1$, that is the 1:1 resonance case. Then the 
Hamiltonian to the potential (1)  is 
\begin{eqnarray}
H&=&\frac{1}{2}\left(p_x^2 + p_y^2 + x^2 + y^2 \right) \nonumber \\ 
&&-  \epsilon \left[ \beta \left(x^4 +y^4 \right) + 2 \alpha x^2 y^2
\right] = h ~~,
\end{eqnarray}
where $p_x, p_y$  are the momenta  per  unit mass conjugate to x, y
and h is the numerical value of H.

 Our aim is to study the various types of periodic orbits, their
stability and the kind of non periodic motion  (regular or chaotic) in
the Hamiltonian  (2) for 
various values of parameters $\alpha $ and $\beta $ using  
the map corresponding to the Hamiltonian (2), as well as numerical
integration. This resonance case  is  also known as the perturbed
elliptic oscillators (Deprit \cite{d1}, Deprit \& Elipe \cite{d2},
 Caranicolas
\& Innanen \cite{ci}, Caranicolas \cite{c2}). The $x-p_x$ Poincare
phase plane 
derived by the map and the numerical integration will be compared in
each case. Of a special interest is the study of the chaotic
motion. We shall try to find an answer to questions such as: 
\begin{enumerate} 
\item Does the map describe in a satisfactory way the chaotic layers  in
the $x-p_x$  plane and, if so, how this behavior evolves by increasing
$\epsilon$ ? 
\item What are the differences, if any, in the  Lyapunov
Characteristic Number (LCN) found by the map and the numerical
integration in the regular and the chaotic area? 
\item Are there any similarities in the invariant spectra derived by
the map and the numerical  integration? 
\end{enumerate}

	The map and the stability conditions of the periodic orbits
are given in Section 2. In the same Section  we compare the $x-p_x$
phase plane found by the map and numerical integration for some of
the main different cases. In Section 3 we compare the LCNs and the
spectra of orbits, derived using the map and numerical
integration. Section 4 is devoted to a discussion and the conclusions
of this work.

\section{Map and  stability  conditions}  

\begin{figure}
\resizebox{\hsize}{!}{\includegraphics*{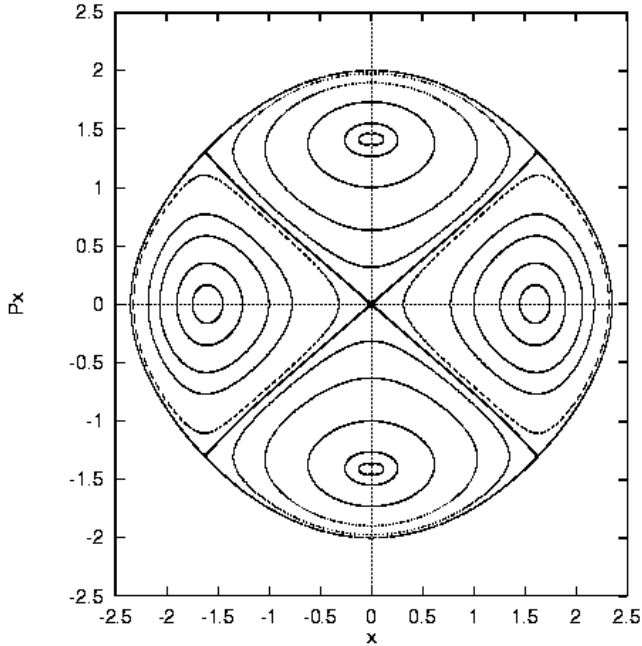}}
\caption{The $x-p_x$ (type A) phase plane derived by numerical
integration. The values of the parameters are  $\alpha=1.2$,
$\beta=0.8$, $\epsilon=\epsilon_{esc}=0.03125$, $h=2$.}
\label{f1a}
\end{figure} 
\begin{figure}
\resizebox{\hsize}{!}{\includegraphics*{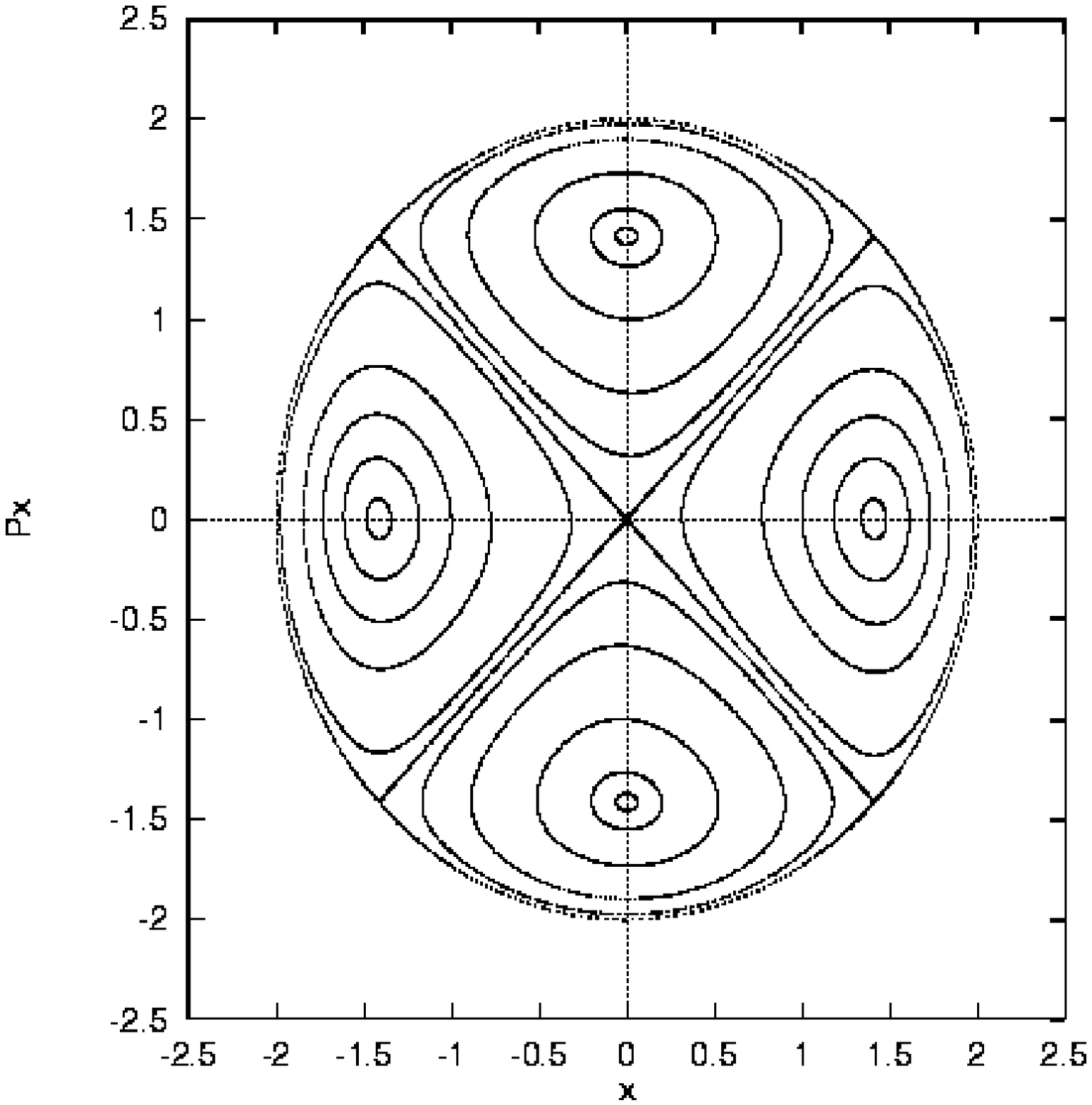}}
\caption{Same as Fig.~\ref{f1a} derived by the map.}
\label{f1b}
\end{figure} 
\begin{figure}
\resizebox{\hsize}{!}{\includegraphics*{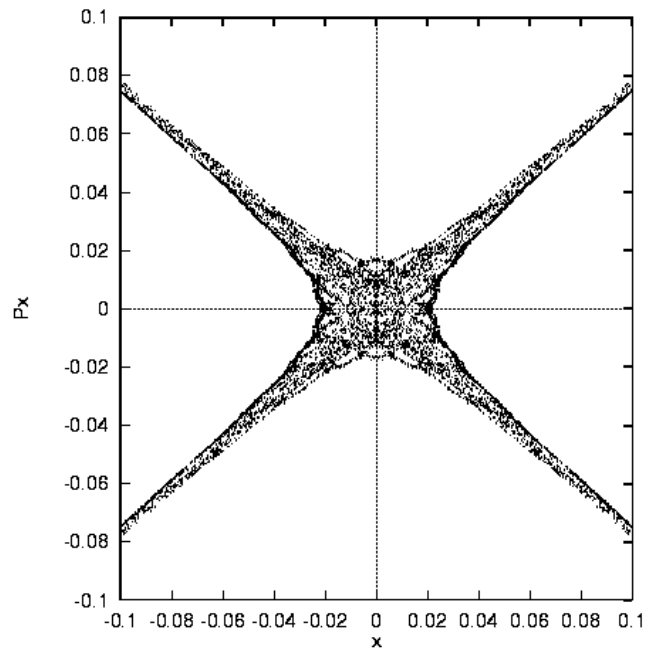}}
\caption{The area near the center of Fig.~\ref{f1a}}
\label{f1c}
\end{figure} 

The averaged Hamiltonian corresponding to  the Hamiltonian (2) reads 
\begin{eqnarray}
<H>&=& \epsilon \alpha J \left(\Lambda - J\right)\left[2- \cos 2
 \phi\right]
 \nonumber \\
&&+ \frac{3}{2}
\epsilon \beta \left[J^2 + \left(\Lambda - J \right)^2\right]
\end{eqnarray}
Following the steps described in Caranicolas (\cite{c1}) we find the map
\begin{eqnarray}
J_{n+1}&=&J_n - 2 \epsilon \alpha J_{n+1} (\Lambda - J_{n+1})\sin 2
\phi_n \nonumber \\ 
\phi_{n+1}&=&\phi_n + \epsilon (2 \alpha - 3 \beta)(\Lambda - 2 J_{n+1})
\\
&& - \epsilon \alpha (\Lambda - 2 J_{n+1}) \cos 2 \phi_n \nonumber
\end{eqnarray}
where  $\Lambda =h$. The map describes the motion in the $J- \phi$
plane and  we return to $x, p_x$ variables  
through  $x=(2J)^{1/2} \cos \phi , p_x=-(2J)^{1/2} \sin \phi$.
The fixed points of (4) are at
\begin{eqnarray}
(i)  \ \ \    J&=&0, \ \ \ \ \ \ \ \ for \ \ \ any \ \ \ \phi \nonumber
\\
(ii) \ \ \    J&=& \Lambda /2 , \ \ \ \ \phi = 0, \pi    \\    
(iii) \ \ \   J&=& \Lambda /2 , \ \ \ \ \phi = \pm \pi /2 ~~. \nonumber 
\end{eqnarray}

There are  three distinguished  case of the $x-p_x$   phase plane
portrait in the Hamiltonian (2). Type A, when both fixed  points (i)
and (ii) are stable. Type B when points (i) are stable while fixed
points (ii) are unstable. In type C phase plane, fixed points (i)  and
(ii) are unstable and stable respectively. Applying the stability
conditions (see Lichtenberg \& Lieberman \cite{ll}) we find after some
straightforward calculations: Type A  phase plane  
appears when     $\alpha > \beta > \alpha/3$  $(\alpha >0 , \beta
>0)$.  Type B appears when $\beta > \alpha $  $(\alpha >0, \beta >0
)$.  We observe the  phase plane of type C when $\alpha > 3 \beta $ $(
\alpha >0, \beta >0)$  or if  $\alpha <0 , \beta >0$. 

In all three cases there is, for a fixed value of  the energy h, a
value of the perturbation strenght $\epsilon_{\mathrm{esc}}$,  such as
for 
$\epsilon > \epsilon_{\mathrm{esc}}$  curves of zero velocity  $h-V=0$
open and 
the test particle is free to escape. We do not consider cases where
the curves of zero velocity are always closed that is cases where the
Hamiltonian (1) has no $\epsilon_{\mathrm{esc}}$. The value of
$\epsilon_{\mathrm{esc}}$ can be found using the method described in
Caranicolas \& Varvoglis  (\cite{cvar}). For  the type A phase  
plane we find 
\begin{equation}
\epsilon_{\mathrm{esc}}=1/[8h(\alpha + \beta )] ~~,
\end{equation}
while in the cases B and C $\epsilon_{\mathrm{esc}}$ is given by the
formula  
\begin{equation} 
\epsilon_{\mathrm{esc}}=1/[16h \beta ] ~~.
\end{equation}

Let us now go to see the three different types of the phase
plane produced by the Hamiltonian system (2). In all numerical
calculation we use the value $h=2$. Fig.~\ref{f1a} shows the type A
$x-p_x$  phase plane derived by numerical integration. The values of
$\alpha , \beta $ are 1.2, 0.8 respectively while  $\epsilon =
\epsilon_{\mathrm{esc}}=0.03125$. The motion is everywhere regular
except near 
the hyperbolic point in the center and in  a thin strip along the
separatrix.(see Fig.~\ref{f1c}). Fig.~\ref{f1b} is the corresponding
figure  produced by the 
map. As one can see the agreement is good. One significant difference
is that 
the map is inadequate to produce the chaotic layer seen in
Fig.~\ref{f1c}.   
Also note that $x_{max}$ in Fig.~\ref{f1a} is greater than 2 while in
Fig.~\ref{f1b} is smaller.

\begin{figure}
\resizebox{\hsize}{!}{\includegraphics*{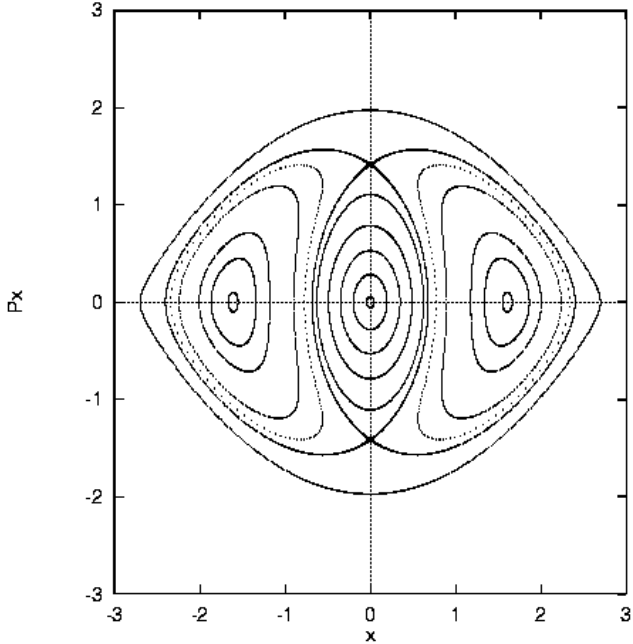}}
\caption{The $x-p_x$ (type B) phase plane derived by numerical
integration. The values of the parameters are  $\alpha=0.2$, 
$\beta=0.25$, $\epsilon=\epsilon_{esc}=0.125$, $h=2$.}
\label{f2a}
\end{figure} 
\begin{figure}
\resizebox{\hsize}{!}{\includegraphics*{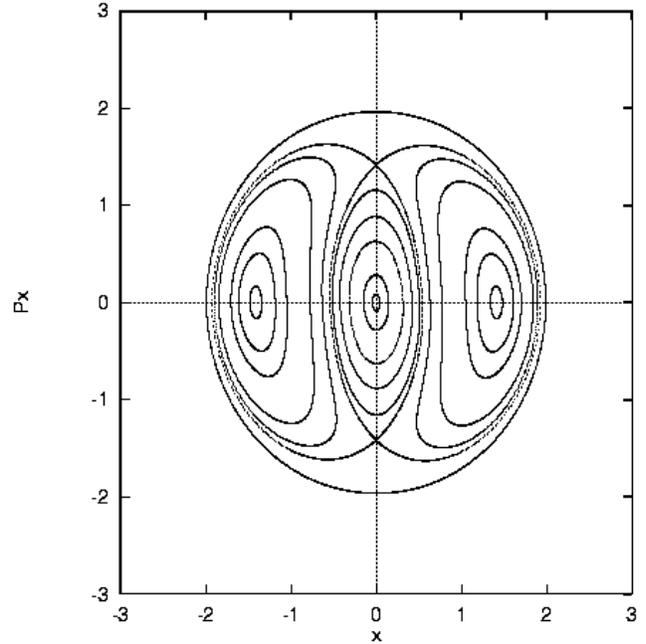}}
\caption{Same as Fig.\ref{f2a} derived by the map.}
\label{f2b}
\end{figure} 
Fig.~\ref{f2a} and Fig.~\ref{f2b} show the type B phase plane derived
using numerical 
integration and the map respectively. The values of the parameters are
$\alpha=0.20$, $\beta=0.25$, while  $\epsilon =
\epsilon_{\mathrm{esc}}= 0.125$.  
The results are similar to those observed in Figs.~\ref{f1a} and
\ref{f1b}. Again the map 
describes well the real phase plane except in a small chaotic region
near the two hyperbolic fixed points. Thus, our numerical calculations
suggest that, the map (4) describes well the   properties of motion in
the Hamiltonian (2) up to the largest perturbation, that is $\epsilon
= \epsilon_{\mathrm{esc}}$   but it is insufficient to  described the
small 
chaotic region observed  when using numerical integration. 

\begin{figure}
\resizebox{\hsize}{!}{\includegraphics*{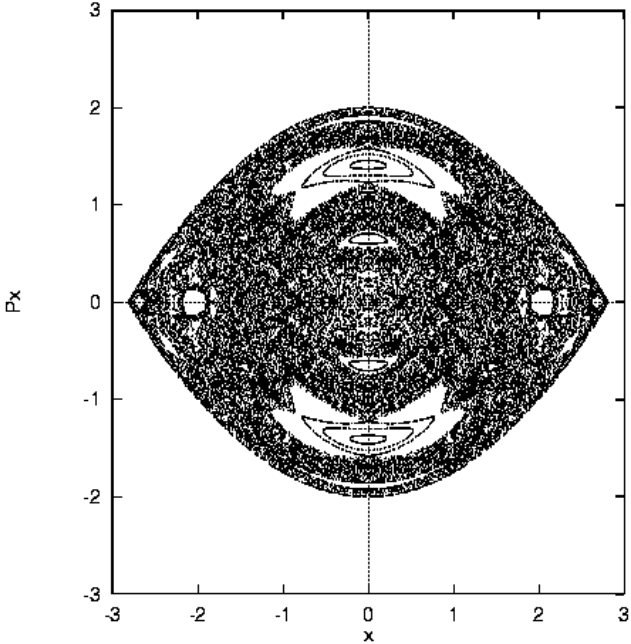}}
\caption{The $x-p_x$ (type C) phase plane derived by numerical
integration. The values of the parameters are  $\alpha=-1.2$, 
$\beta=0.2$, $\epsilon=\epsilon_{esc}=0.15625$, $h=2$.}
\label{f3a}
\end{figure}

\begin{figure}
\resizebox{\hsize}{!}{\includegraphics*{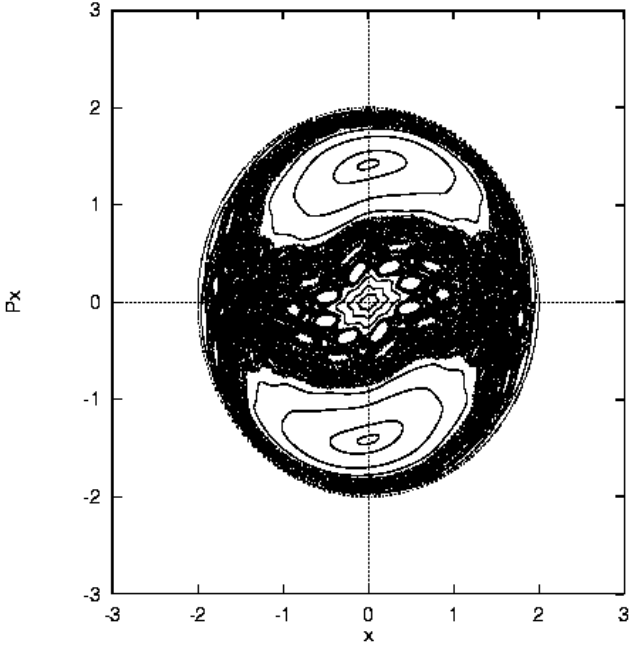}}
\caption{Same as Fig.\ref{f3a} derived by the map.}
\label{f3b}
\end{figure}

	Type C phase plane  is shown in Figs.~\ref{f3a} and \ref{f3b}.
Fig.~\ref{f3a}
comes from numerical integration  while Fig.~\ref{f3b}  was derived using
the map.  
The values of the parameters are $\alpha =-1.20 , \beta =0.20$, while
$\epsilon =\epsilon_{\mathrm{esc}}= 0.15625$. This value of $\epsilon$
was chosen to maximize the chaotic effects. The most important
characteristic, observed in both Figs.~\ref{f3a} and \ref{f3b} is the
large unified 
chaotic region. In contrast with the two previous cases A and B, here
the map reproduces satisfactorily the chaotic sea found by numerical
integration. On the other hand, it   is evident that the map describes
qualitatively in a satisfactory  way the areas of regular motion
around the elliptic 
points in the $p_x$ axis. Some differences are observed in the area
near the center between the two patterns. Another important
characteristic, observed in case  C, is that  the chaotic areas are
large even when $ \epsilon < \epsilon_{\mathrm{esc}}$. Results, not
shown here, 
suggest that considerable chaotic areas  are observed  in the phase
plane derived using the map, when $\epsilon \geq 0.08$. Therefore we
must admit  that our numerical experiments show that, in the case when
the Hamiltonian system (2) has small chaotic regions,  the map is
inadequate to produce them but it describes them satisfactorily  when
they are large.  

\section{Dynamical spectra}

In  this Section we shall study the spectra of the Hamiltonian  system
(2) using  numerical integration and the map (4).Before doing this, it
is necessary  to remember some useful notions and definitions. The
``stretching number'' $\alpha_i$ (see Voglis \& Contopoulos \cite{vogcon},
Contopoulos at al. \cite{cong}, Contopoulos \& Voglis \cite{conv}) is
defined as   
\begin{equation}
\alpha_i=\ln \left| \frac{\xi_{i+1}}{\xi_i} \right|~~,
\end{equation}
where  $\xi_{i+1}$ is the next image on the Poincare phase plane of an
infinitesimal deviation $\xi_i$  
between two nearby  orbits. The spectrum of the stretching numbers is
their distribution function
\begin{equation}
S(\alpha_i)=\frac{\Delta N (\alpha)}{N d \alpha}~~,
\end{equation}
where $\Delta Π(\alpha )$ is the number of stretching numbers  in
the interval  $( \alpha , \alpha + d \alpha )$  after N  
iterations.The maximal Lyapunov characteristic number can be written as
\begin{equation}
LCN=\int \limits_{-\infty}^{\infty} \alpha S( \alpha ) d \alpha~~,
\end{equation}
in other words, the LCN is the average value
 of the stretching number $\alpha$. 

Today we know that the distribution  of the successive stretching
numbers forms a spectrum which is invariant with respect to (i) the
initial conditions along an orbit and the direction of the deviation
from this orbit and (ii) the initial conditions of the orbits
belonging to the same chaotic region.  

In what follows we shall give the spectra $S ( \alpha )$ of  orbits of
the Hamiltonian system (2) derived using the map (4) as well as
numerical integration. In the case of the Hamiltonian where t is a
continuous quantity ( that is in the numerical integration of the
equations of motion) we use for the derivation of the stretching
numbers and spectra the method used by Contopoulos et al. (1995). All
calculations correspond to the case C 
because  the results are much more
interesting.     
\begin{figure}
\resizebox{\hsize}{!}{\rotatebox{270}{\includegraphics*{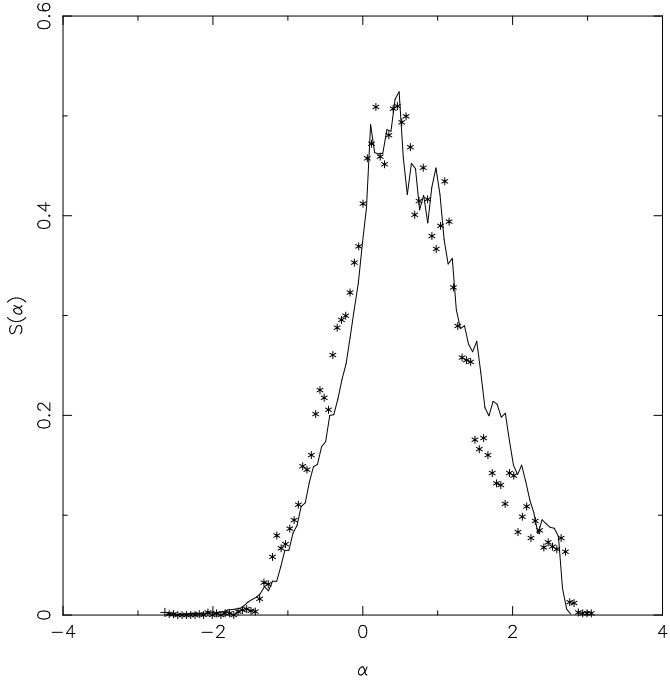}}}
\caption{The spectrum $S(\alpha)$ of two chaotic  orbits derived
using numerical integration. The values of the parameters are as in
Fig.~\ref{f3a}. Initial conditions $x=1.3$, $y=p_x=0$ and $x=1.1$, $y=0$,
$p_x=0.2$. The value of $p_y>0$ is found from the energy integral.}
\label{f4}
\end{figure} 
\begin{figure}
\resizebox{\hsize}{!}{\rotatebox{270}{\includegraphics*{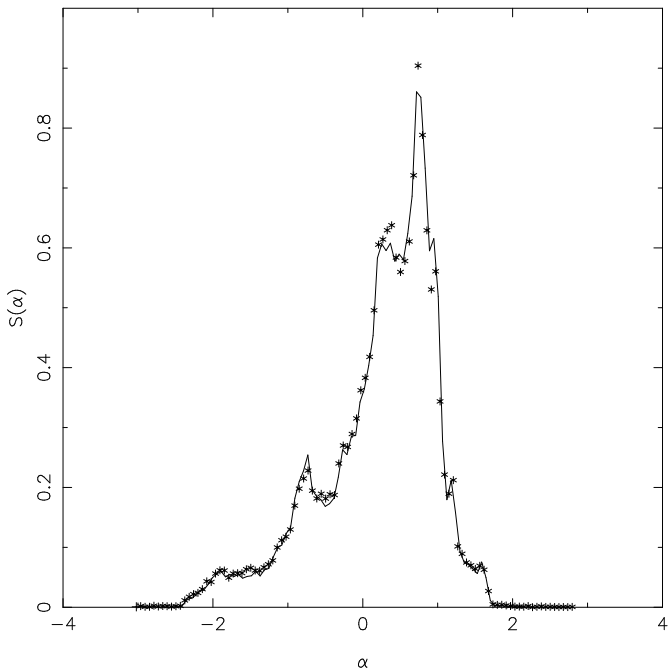}}}
\caption{Same as Fig.~\ref{f4} derived using the map.}
\label{f5}
\end{figure}
\begin{figure}
\resizebox{\hsize}{!}{\includegraphics*{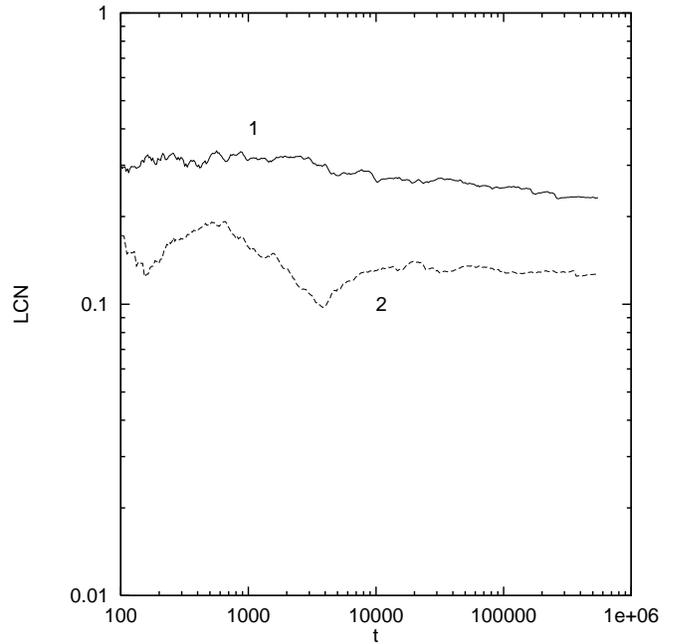}}
\caption{ LCNs  for the same chaotic orbit given by the map 1  and
numerical integration 2. Initial conditions $x=1.3$, $y=p_x=0$.} 
\label{f6}
\end{figure} 

Fig.~\ref{f4} shows the spectra of two orbits (the first with solid
line the other with dots) derived using numerical integration. The
orbits were started in the  chaotic region 
with different initial conditions. It is evident that the two
spectra are close to each other. Fig.~\ref{f5} shows the spectra of
the same two orbits found using the map. Again the two spectra
are close to each other. The spectra in both cases were
calculated for $N=10^6$ periods. As one observes the pattern
shown in both Figs.~\ref{f4} and \ref{f5} have the  characteristics of
the 
spectrum of  orbits belonging to a chaotic region. Fig.~\ref{f6}
shows the LCNs  for two chaotic orbits. Number 1 was derived 
using the map, while number 2 was found using  numerical integration.
 As one can see
the mean exponential divergence of the two
nearby chaotic trajectories, described by  the map, is larger
than that given by numerical  integration. Nevertheless the two curves
are qualitatively similar.    

\begin{figure}
\resizebox{\hsize}{!}{\rotatebox{270}{\includegraphics*{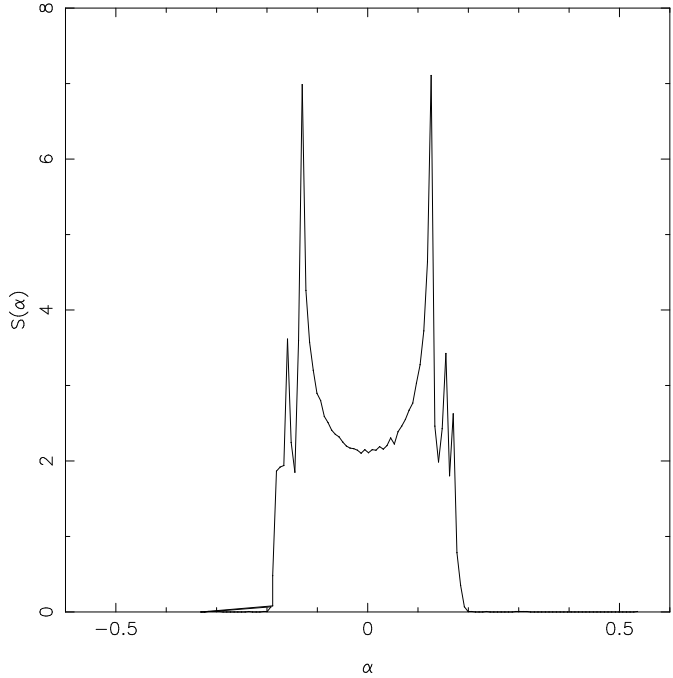}}}
\caption{The spectrum  $S(\alpha)$ of  a regular orbit derived using
numerical integration. The values of the parameters are as in
Fig.~\ref{f3a}. Initial conditions $x=y=0$, $p_x=1.55$. The value of
$p_y>0$ is found from the energy integral.} 
\label{f7}
\end{figure}

\begin{figure}
\resizebox{\hsize}{!}{\rotatebox{270}{\includegraphics*{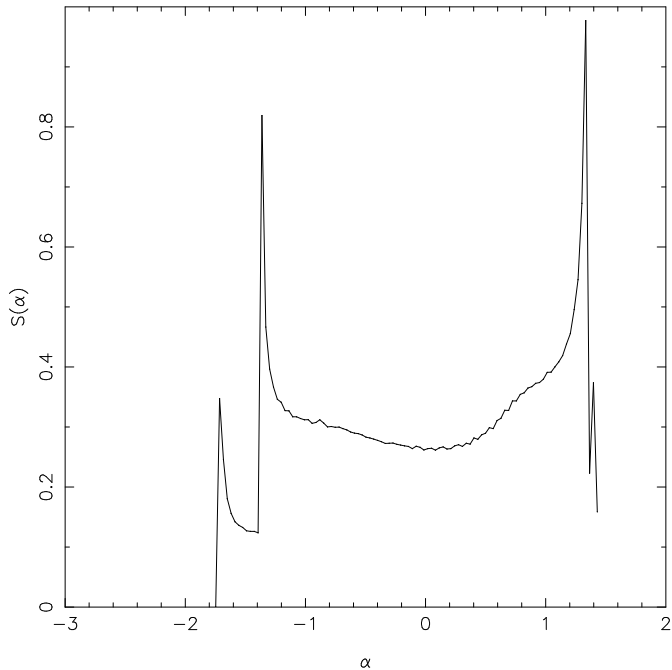}}}
\caption{Same as Fig.~\ref{f7} derived using the map.} 
\label{f8}
\end{figure}

\begin{figure}
\resizebox{\hsize}{!}{\includegraphics*{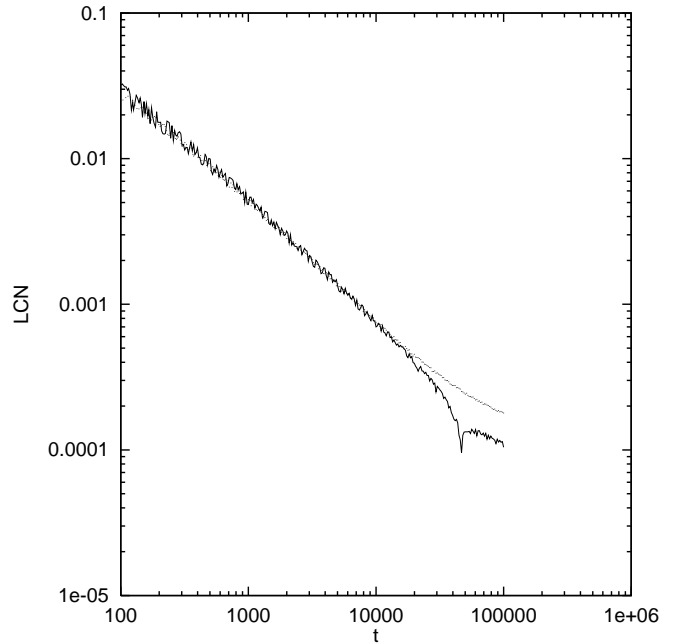}}
\caption{ LCNs  for the same regular orbit given by the map (dots)
and numerical integration (solid line). Initial conditions as in
Fig~\ref{f7}. } 
\label{f9}
\end{figure}

In Figs.~\ref{f7} and \ref{f8} we give the spectra of  the same regular
orbit derived by numerical integration and the map respectively. This
is an orbit with initial conditions near the periodic orbit $ x=y=0,
p_x=\Lambda^{1/2} $. The spectra  were calculated for $N=10^6$ periods. As
one can see the agreement between the two spectra is good. 
From other
cases we know that the agreement is much better if we calculate an
orbit for $N=10^7$, or $N=10^8$ periods. 
Both are ``U-shaped'', with two large and two small peaks. Such spectra are
characteristic of quasi-periodic orbits, starting close to a stable
periodic orbit (see Patsis et al. \cite{panos}). 
Fig.~\ref{f9} shows the LCNs
for the orbit of Fig.~\ref{f7} derived  using a map (dots) and numerical
integration (solid line). Again one can see that the map describes
well the qualitative properties of regular orbits.

Let us now come to the symmetry of the spectra. It was observed that
the spectra of ordered orbits, starting close to a stable periodic
orbit, are almost symmetric with  respect to the $\alpha=0$ axis,
while, when we go far from the stable periodic orbit they become
asymmetric. In order to give an estimate  between closeness
and symmetry we have made  extensive numerical experiments near the
exact periodic orbit $x=0$, $p_x=h^{1/2}=\sqrt{2}$.

\begin{figure}
\resizebox{\hsize}{!}{\rotatebox{270}{\includegraphics*{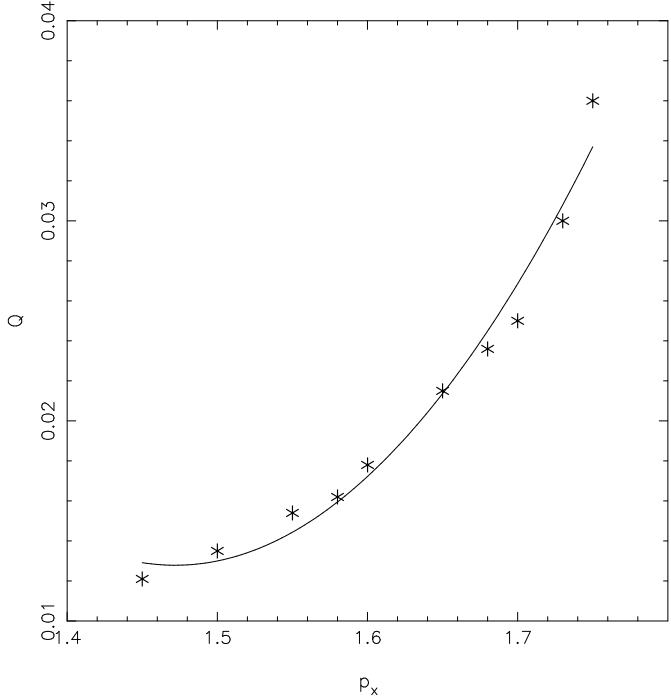}}}
\caption{$Q$ vs $p_x$ for the map in case C, when $\epsilon=0.02$} 
\label{f10}
\end{figure}

\begin{figure}
\resizebox{\hsize}{!}{\rotatebox{270}{\includegraphics*{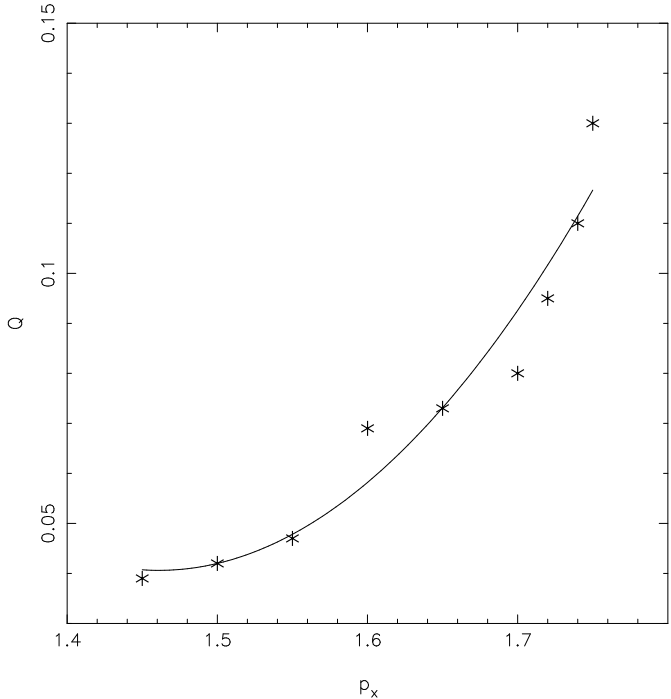}}}
\caption{Same as Fig.~\ref{f10} for the Hamiltonian in case C 
when $\epsilon=0.1$.} 
\label{f11}
\end{figure}

We define as ``symmetry factor'' the quantity
\begin{equation}
Q=\frac{1}{N}\sum_{i=1}^N | S(\alpha_i) - S(-\alpha_i) |
\end{equation}
As one can see, $Q=0$ corresponds to a perfectly symmetric spectrum,
while $Q=1$ to a totally asymmetric with all non zero values of
$S(\alpha)$ 
to belong to either positive or negative $\alpha$. The results $Q$ vs
$p_x$ for the map, in the case C, when $\epsilon=0.02$, are shown in
Fig.~\ref{f10}. Dots correspond to values given by the numerical experiments,
while the solid line corresponds to the best fit
\begin{equation}
Q~=~\mathrm{a} p_x^2 + b p_x +c~~~,
\end{equation}
where $\mathrm{a}=0.2704$, $b=-0.7960$ and $c=0.5986$.
Fig.~\ref{f11} is the same as Fig.~\ref{f10}, for the Hamiltonian in
the case C  when
$\epsilon=0.1$.
The best fit is now with $\mathrm{a}=0.9100$, $b=-2.659$ and $c=1.983$.
As we can see we have a second order polynomial growth of the
asymmetry of the spectrum as
we move far from the stable periodic orbit. The value of N in all
cases was $10^6$ iterations.

\section{Discussion}

	In this paper we have studied the regular and chaotic motion
in the Hamiltonian  system (2) using  a  map, derived from the averaged
Hamiltonian and numerical integration. Note that Hamiltonian (2) is
integrable when $ \alpha = \beta$ and when  $ \alpha = 3
\beta$. Depending on the choice of the parameters $ \alpha, \beta$
this Hamiltonian displays three different types of $x-p_x$ phase plane
named as types A,B,C. In the case of the two first types of phase
plane, the map describes in a  very satisfactory way the real
properties of orbits up to $\epsilon =
\epsilon_{\mathrm{esc}}$. Extensive 
numerical calculations have shown that the map does not produce any
chaotic regions in the cases A and B. Therefore, one concludes that
the map fails to describe the small chaotic regions  near the
separatrix found by numerical  integration. Here we must note that at
the beginning we had started with smaller values of $\epsilon$, but we
observed that the map persisted not to produce chaos up to
$\epsilon=\epsilon_{\mathrm{esc}}$. This explains why
$\epsilon=\epsilon_{\mathrm{esc}}$ was chosen.


	The situation is quite different in the case of the phase
plane of type C. Here the system  has large chaotic regions, increasing
when $\epsilon $ approaches $\epsilon_{\mathrm{esc}}$ and the map
describes 
them satisfactorily. Moreover, the comparison of the spectra  and LCNs
of orbits found using the map and numerical integration shows that one
can trust the map. This is very important because the map is at least
10-20 times faster than numerical integration and one can make faster
all  the time consuming calculations. Note that in all cases the
number of periods in the calculation of the spectra using the  map or
numerical integration was the same in order to be able to compare the
corresponding results. 

We have also investigated the symmetry of the spectra of ordered
orbits. It was found that the quantity $Q$ increases, with a second
order polynomial dependence, as we are moving far from the
stable periodic orbit.

Finally the authors would like to make clear that (i) the proposed map
has some qualitative similarities with the Poincar\'e map of the
original Hamiltonian system (2), although the numerical differences may
be important and (ii) the results
of this work correspond to the particular Hamiltonian system (2) and
for the resonance case 1:1. Interesting results,  on the spectra of
orbits, in galactic Hamiltonians made up of harmonic oscillators in the
4:3 resonance,  have been given by Contopoulos et al. (\cite{cong}). Also
spectra of orbits have been studied, in the standard map, by Voglis \&
Contopoulos (\cite{vogcon}), Contopoulos et al. (\cite{conve}). This paper was
focused 
on the comparison of the results (structure of the phase plane and
spectra of orbits)  given by numerical integration and the map.

\begin{acknowledgements}
The authors would like to thank the referee Prof. G. Contopoulos for
valuable suggestions and comments that helped to significantly
improve the paper.
\end{acknowledgements}

\end{document}